\documentclass[aps,prb,reprint,groupedaddress]{revtex4-2}

\usepackage{graphicx}
\usepackage{amsmath}
\usepackage{upgreek}

\begin{document}

\title{Ultrafast and Anisotropic Exciton Transport in Layered NbOI$_2$}

\author{Salman Ahsanullah}
\author{Neema Rafizadeh}
\author{Hui Zhao}
\email{huizhao@ku.edu}
\affiliation{Department of Physics and Astronomy, The University of Kansas, Lawrence, Kansas 66045, USA}

\begin{abstract}
Ferroelectric layered semiconductors provide a unique platform for exploring coupled excitonic and electronic phenomena in reduced dimensions. Here, exciton transport in layered NbOI$_2$ is investigated using transient absorption microscopy. Ultrafast and strongly anisotropic exciton diffusion is observed. The diffusion coefficient of about 160 cm$^2$ s$^{-1}$ along the nonpolar crystal axis is among the highest reported for van der Waals semiconductors. The transport exhibits an approximately twofold in-plane anisotropy between the polar and nonpolar directions, reflecting the intrinsic low-symmetry crystal structure of NbOI$_2$. The environmental stability of NbOI$_2$ is further investigated, and fabrication and encapsulation conditions required to achieve long-term sample stability are identified. These results establish NbOI$_2$ as a promising platform for studying anisotropic exciton transport and highlight the critical role of interface protection in air-sensitive van der Waals materials.

\end{abstract}


\maketitle

\section{Introduction}

Layered niobium oxide dihalides NbOX$_2$ (X = Cl, Br, I) have recently emerged as a distinctive class of van der Waals materials that possess both semiconducting electronic behaviors and intrinsic in-plane ferroelectricity \cite{np181170,nh41113,np19142,nc164987,small212407729}. Among this family, NbOI$_2$ has attracted particular attention as a prototypical in-plane ferroelectric semiconductor with novel optical properties. Its monoclinic crystal structure contains Nb atoms displaced from the centers of [NbO$_2$X$_4$] octahedra, generating a robust in-plane polarization that persists at room temperature, which lowers the lattice symmetry \cite{am332101505}. This combination of broken inversion symmetry and semiconducting behavior enables electrically switchable rectification \cite{apl119033103}, antiferroelectric ordering \cite{nh41113,acsnano177170,nc145911}, and a variety of device concepts based on the polarization-controlled electronic responses. 

The reduced crystal symmetry of NbOI$_2$ manifests as pronounced in-plane anisotropy across multiple types of physical properties \cite{nanotechnology33275701,jap134085105,nc131884}. At the mechanical level, significant directional dependence of Young’s modulus has been measured, reflecting anisotropic bonding associated with the Peierls-distorted lattice \cite{apl123051905}. Lattice vibrational properties also exhibit strong anisotropy: Polarization-resolved Raman and infrared spectroscopic studies have revealed directionally dependent phonon modes and lattice dynamics \cite{am332101505}. The linear optical absorption also shows pronounced anisotropy \cite{am332101505,nl26401}. Furthermore, NbOI$_2$ exhibits exceptionally strong and anisotropic nonlinear optical responses. Giant and electrically tunable second-harmonic generation, as well as higher-order harmonic processes, have been reported with efficiencies exceeding those of many other two-dimensional (2D) materials \cite{np16644,afm342308207,nl243413,afm2025e16389,apr20252500188,nc145911}. The broken inversion symmetry also gives rise to substantial electromechanical and electro-optic coupling. Large piezoelectric coefficients and pronounced linear electro-optic and elasto-optic responses have been demonstrated in both thin flakes and bulk crystals of NbOI$_2$ \cite{nc131884,b110L100101,ss42300125}, highlighting strong coupling between polarization, strain, and built-in electric fields. Furthermore, anisotropic in-plane electrical conductance and photoresponsivity have been revealed \cite{am332101505,am372500049}.

Despite these extensive efforts, little is known about the influence of the low-symmetry ferroelectric lattice on in-plane photocarrier transport. Although steady-state electrical measurements have revealed anisotropic conductance \cite{am332101505,am372500049}, they do not directly probe the transport of photocarriers or excitons on ultrafast time scales. Here, we report transient absorption microscopy measurements of exciton diffusion in NbOI$_2$ crystals at room temperature. We observe a remarkably large exciton diffusion coefficient together with pronounced in-plane anisotropy, directly reflecting the underlying ferroelectric lattice symmetry. The diffusion coefficient reaches values among the highest reported for 2D semiconductors, revealing the unusual exciton transport properties of NbOI$_2$. We also systematically investigate the environmental stability of NbOI$_2$ under different fabrication and encapsulation conditions. Samples prepared under optimized conditions show no visible degradation after long-term ambient exposure and repeated laser measurements. These findings establish NbOI$_2$ as a distinctive platform for exploring exciton transport in low-symmetry ferroelectric materials while identifying a practical approach for preserving this air-sensitive material.

\section{Sample Fabrication and Stability}

Optical investigations, particularly spatially and temporally resolved laser spectroscopy, require samples that remain stable over extended periods of repeated measurements. NbOI$_2$ has previously been reported to degrade under ambient conditions, although the reported degree of stability varies considerably across the literature \cite{acsami169051,apl128091904,am372500049,am372507168}. We therefore compared several fabrication and encapsulation procedures to establish a reliable method for preparing NbOI$_2$ samples for transient absorption microscopy.

\begin{figure*}[t]
  \centering
  \includegraphics[width=0.8\textwidth]{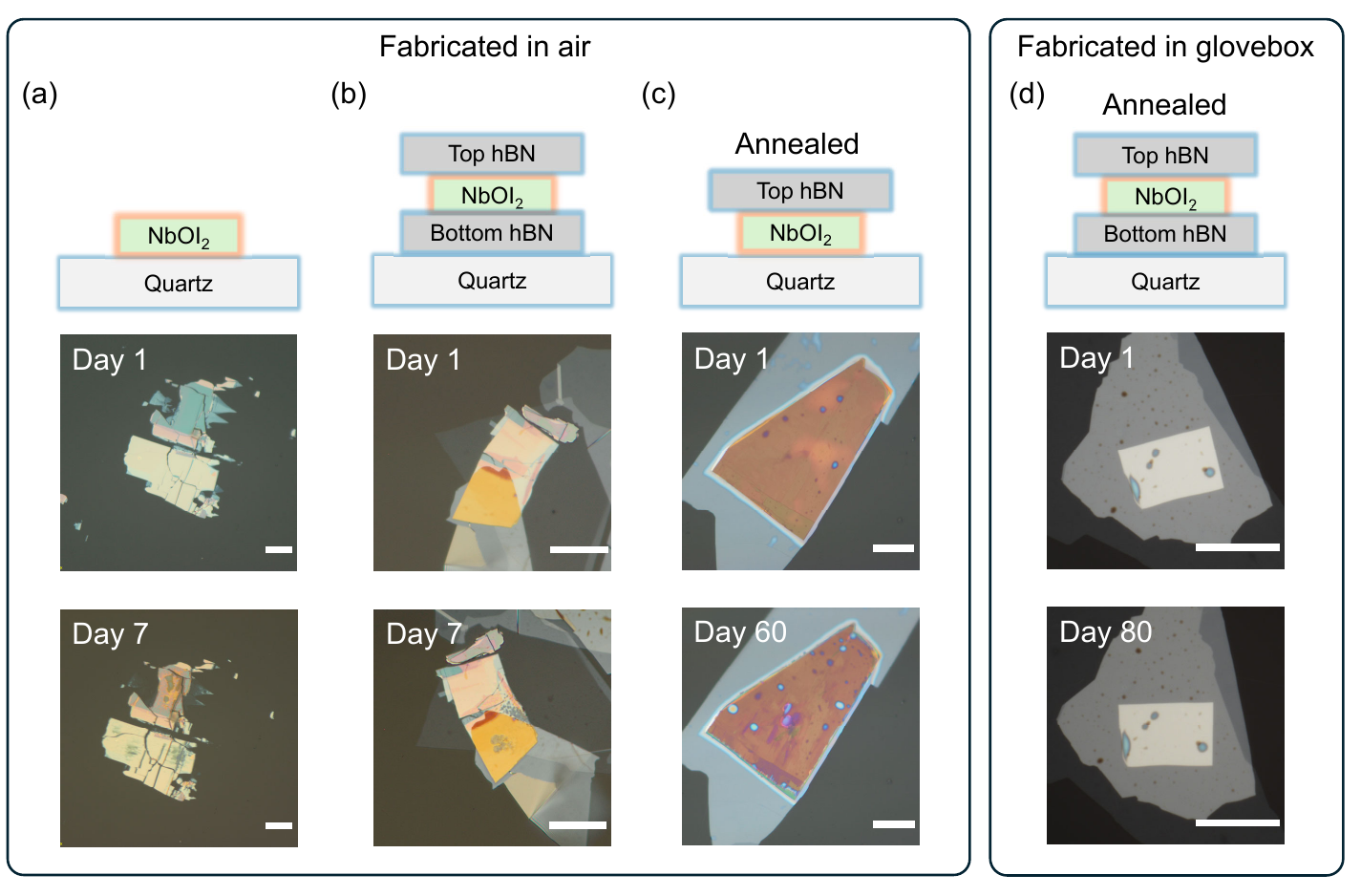}
  \caption{Optical microscope images showing the stability of NbOI$_2$ samples prepared under different fabrication conditions. (a)--(c) Samples fabricated under ambient conditions: (a) unencapsulated NbOI$_2$, (b) NbOI$_2$ fully encapsulated between top and bottom hBN layers without thermal annealing, and (c) NbOI$_2$ encapsulated by a top hBN layer and thermally annealed. (d) NbOI$_2$ fabricated in an Ar-filled glovebox, fully encapsulated between hBN layers, and thermally annealed for 3~h. No visible degradation is observed after 80 days of ambient exposure and repeated laser measurements. Scale bars in (a)--(d) are 20~$\upmu$m.}
  \label{fig:degradation}
\end{figure*}

Figure~\ref{fig:degradation} compares the stability of NbOI$_2$ samples prepared under different conditions. Flakes mechanically exfoliated from bulk crystals and transferred onto quartz under ambient conditions without encapsulation deteriorate within a few days, even with less than one hour of cumulative laser exposure [Figure~\ref{fig:degradation}(a)]. This observation confirms that unprotected NbOI$_2$ is susceptible to degradation under ambient conditions and laser illumination, consistent with previous reports attributing its instability to reactive iodine bonds \cite{acsami169051,apl128091904}.

Encapsulation with hexagonal boron nitride (hBN) is commonly used to protect air-sensitive thin materials. However, we find that encapsulation alone does not ensure the long-term stability of NbOI$_2$. As shown in Figure~\ref{fig:degradation}(b), a sample fully encapsulated between top and bottom hBN layers, each several tens of nanometers thick, but not thermally annealed also deteriorates rapidly. This degradation likely results from residual adsorbates and air trapped at the interfaces during fabrication under ambient conditions.

Sample stability improves substantially when thermal annealing is performed after the transfer of each flake. The sample fabricated under ambient conditions with top hBN encapsulation and thermal annealing exhibits noticeable degradation only after several weeks [Figure~\ref{fig:degradation}(c)]. Each annealing step was performed at 150~$^\circ$C for 1~h. We attribute the improved stability to enhanced interlayer contact and the removal of trapped adsorbates and air pockets, thereby reducing interfacial disorder and potential pathways for degradation.

The greatest stability is achieved when the entire fabrication process is performed in an Ar-filled glovebox and combined with full hBN encapsulation and thermal annealing. No visible degradation is observed after 80 days of ambient exposure and repeated laser measurements [Figure~\ref{fig:degradation}(d)]. By contrast, flakes exfoliated in the glovebox but left unencapsulated degrade rapidly, confirming that fabrication in an inert environment alone is insufficient.

These comparisons show that the combination of inert-environment fabrication, full hBN encapsulation, and thermal annealing provides the best long-term stability among the procedures investigated. The sample shown in Figure~\ref{fig:degradation}(d) was therefore used for the transient absorption measurements. The NbOI$_2$ flake is approximately 21~nm thick, as determined by atomic force microscopy.

\section{Transient absorption measurements}

 \begin{figure}[t]
  \centering
  \includegraphics[width=\columnwidth]{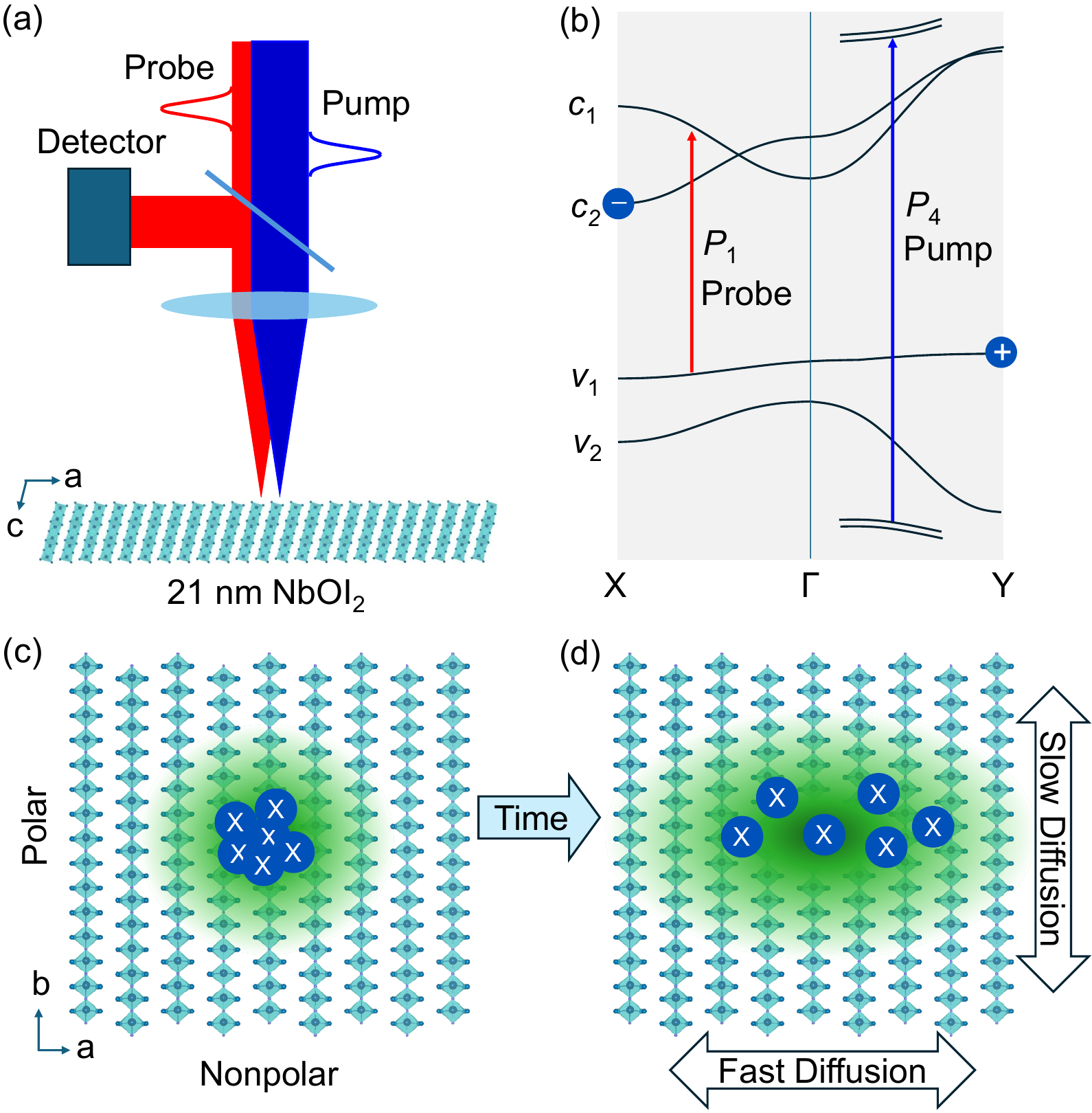}
  \caption{(a) Schematic of the transient absorption microscopy setup, showing tightly focused pump and probe pulses used to excite and detect the NbOI$_2$ sample. (b) Schematic illustration of the NbOI$_2$ band structure and the corresponding pump–probe excitation scheme. (c) Initially, excitons (X) are generated by the pump pulse with a narrow Gaussian distribution. (d) Anisotropic diffusion along the polar and nonpolar directions results in an elongated exciton distribution at later times, reflecting faster diffusion along the nonpolar axis and slower diffusion along the polar axis.}
    \label{fig:scheme}
\end{figure}

Figure~\ref{fig:scheme} summarizes the experimental scheme. In the transient absorption microscopy measurement [Figure~\ref{fig:scheme}(a)], a tightly focused pump pulse excites a NbOI$_2$ flake under normal incidence. Figure~\ref{fig:scheme}(b) illustrates the key features of the NbOI$_2$ band structure, including the two lowest conduction bands ($c_1$ and $c_2$) and two highest valence bands ($v_1$ and $v_2$), as well as the excitonic transitions $P_1$ and $P_4$, established by angle-resolved photoemission spectroscopy, optical spectroscopy, and density functional theory \cite{np16644}. As an indirect semiconductor, the conduction-band minimum and valence-band maximum are located at the X and Y points, respectively. In our experiment, a 3.02~eV pump excites the system via the $P_4$ transition, after which electrons and holes relax to the X and Y valleys, respectively. Given the large exciton binding energy in this material family, which is on the order of hundreds of meV \cite{n61353,jpcl159191}, the photoexcited population is expected to be dominated by excitons formed from these carriers.

 \begin{figure*}[t]
  \centering
  \includegraphics[width=\textwidth]{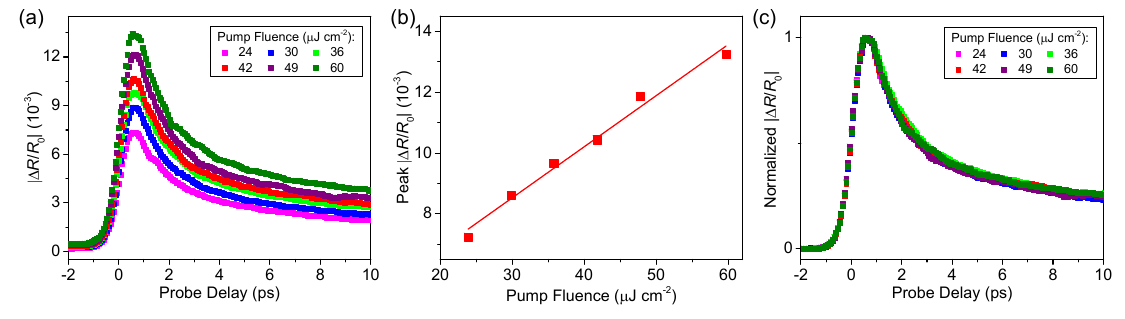}
  \caption{(a) Differential reflectance of NbOI$_2$ as a function of probe delay for various pump fluences, measured with a 3.02 eV pump and a 2.25 eV probe.
(b) Peak differential reflectance as a function of pump fluence, showing a linear dependence.
(c) Normalized differential reflectance traces demonstrating that the decay dynamics are independent of pump fluence.}
    \label{fig:DR}
\end{figure*}

Figure~\ref{fig:scheme}(c) illustrates the crystal structure of the sample surface (the $a$--$b$ plane). The displacement of Nb atoms along the $b$ axis induces a spontaneous polarization along this direction, which we define as the polar axis. The orthogonal in-plane direction (the $a$ axis) is therefore referred to as the nonpolar direction. Owing to the Gaussian intensity profile of the pump pulse, excitons (X) are initially generated with a Gaussian spatial distribution,
\begin{equation}
n(x,0) = n_0 \exp\left(-\frac{x^2}{2\sigma_0^2}\right),
\end{equation}
where $\sigma_0^2$ is the initial variance. The excitons subsequently diffuse away from the excitation spot. Along the nonpolar direction, the density profile evolves as
\begin{equation}
n(x,t) = n_0 A(t) \exp\left(-\frac{x^2}{2\sigma_{\mathrm{np}}^2(t)}\right),
\label{eq:n}
\end{equation}
according to the classical diffusion-recombination model \cite{arpc711,afm271604509}. Here, $A(t)$ accounts for the temporal decay of the peak amplitude and the variance increases linearly with time,
\begin{equation}
\sigma_{\mathrm{np}}^2(t) = \sigma_0^2 + 2D_{\mathrm{np}} t,
\label{eq:sigma2_np}
\end{equation}
with $D_{\mathrm{np}}$ denoting the diffusion coefficient along the nonpolar direction. Similarly, along the polar direction, the variance evolves as
\begin{equation}
\sigma_{\mathrm{p}}^2(t) = \sigma_0^2 + 2D_{\mathrm{p}} t.
\label{eq:sigma2_p}
\end{equation}
Anisotropic exciton transport therefore leads to different expansion rates of the spatial profile along the two orthogonal directions, as schematically illustrated in Figure~\ref{fig:scheme}(d).

Transient absorption microscopy was performed at room temperature under ambient conditions using a home-built system that integrates an ultrafast pump–probe setup with an optical microscope. The light source consists of a Ti:sapphire oscillator (Spectra-Physics Mai Tai) and an optical parametric oscillator (Spectra-Physics Inspire). The Mai Tai produces 100 fs pulses at 1.51 eV with an 80 MHz repetition rate. Its output is directed into the Inspire, where frequency doubling generates 3.02 eV pulses to drive parametric oscillation. The signal output, tunable across the visible range, is used as the probe pulse, while the residual 3.02 eV beam serves as the pump. The pump and probe beams are combined using a beamsplitter and focused onto the sample through an objective lens, producing spot sizes of approximately 1–2~$\upmu$m. The reflected probe beam is collected by the same objective and detected by a silicon photodiode. Optical filters are placed in front of the detector to suppress reflected and scattered pump light. The photodiode output is measured using a lock-in amplifier referenced to a mechanical chopper that modulates the pump beam at $\sim$3 kHz. This configuration enables measurement of the differential reflectance, defined as $\Delta R / R_0 = (R - R_0) / R_0$, where $R$ and $R_0$ denote the reflectance of the sample in the presence and absence of the pump pulse, respectively. Spatial resolution is achieved by scanning the probe spot across the fixed pump spot. This is implemented by tilting a mirror mounted on a motorized mirror mount in the probe beam path, thereby varying the incident angle at the objective lens. The scan distance is calibrated by imaging a reference structure with known dimensions. For spatiotemporal measurements, the probe position is scanned spatially, and at each position the pump–probe delay is varied by adjusting the optical path length of the probe arm. This delay is controlled by translating a retroreflector mounted on a motorized linear stage.

Excitons are injected by the 3.02 eV pump pulse, and monitored by a tightly focused, time-delayed probe pulse. In recent work, we measured the transient absorption spectrum of NbOI$_2$ and found that the $P_1$ transition provides an effective probe of exciton dynamics in the time domain \cite{nl26401}. Here, we employ a 2.25~eV probe tuned near the $P_1$ transition \cite{np16644,nl26401} and measure the differential reflectance, $\Delta R/R_0$, as a function of both the pump–probe delay and the spatial separation between the pump and probe spots. This approach allows us to directly resolve the spatiotemporal evolution of the exciton distribution \cite{afm271604509}.

We first verify that the measurements are performed in the linear regime and that the exciton lifetime is independent of exciton density. These conditions are essential for reliably extracting exciton transport properties from the measured signal. We measured differential reflectance over a range of pump fluences, corresponding to different injected exciton densities. Figure~\ref{fig:DR}(a) presents the time-resolved differential reflectance at various pump fluences. The signal is negative; for clarity, its absolute value is plotted. The peak signal, obtained shortly after zero probe delay, scales linearly with pump fluence (and therefore with exciton density) as shown in Figure~\ref{fig:DR}(b). This linear relationship confirms that the signal faithfully tracks the exciton population. Figure~\ref{fig:DR}(c) shows the normalized traces overlap, demonstrating identical temporal dynamics across all excitation densities. The decay rate of the signal is therefore independent of exciton density. Such density-independent recombination ensures that the exciton recombination causes the exciton population to decrease uniformly in space without altering the spatial profile. Consequently, any broadening of the spatial distribution with time arises solely from exciton diffusion.

 \begin{figure*}[t]
  \centering
  \includegraphics[width=\textwidth]{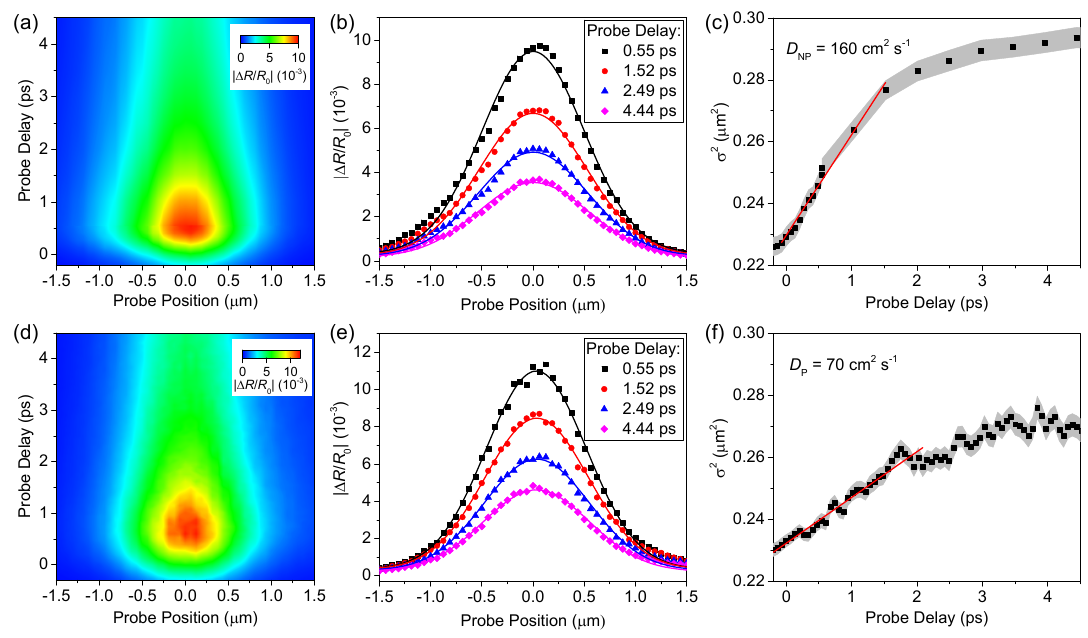}
  \caption{(a) Differential reflectance of NbOI$_2$ at room temperature, measured with a 3.02 eV pump and a 2.25 eV probe. The probe spot is scanned relative to the pump spot along the nonpolar direction, while the probe delay is scanned at each spatial position. (b) Examples of spatial profiles of the differential reflectance at selected probe delays, together with Gaussian fits (curves). (c) Variance of the spatial profile along the nonpolar direction as a function of probe delay. A linear fit yields a diffusion coefficient of 160~cm$^2$~s$^{-1}$. (d–f) Corresponding measurements obtained by scanning the probe spot along the polar direction, yielding an exciton diffusion coefficient of 70~cm$^2$~s$^{-1}$.}
    \label{fig:diffusion}
\end{figure*}

Figure~\ref{fig:diffusion} shows the results of the spatially and temporally resolved transient absorption microscopy measurements with a pump fluence of 36~$\upmu$J cm$^{-2}$. We first investigated exciton diffusion along the nonpolar direction by scanning the probe spot relative to the pump spot along the nonpolar axis of the NbOI$_2$ sample. The orientations of the polar and nonpolar crystal axes of this sample were determined by polarization-dependent absorption measurements in our previous work \cite{nl26401}. Panel (a) presents the spatiotemporally resolved differential reflectance. At each probe delay, the spatial profile can be well fitted by a Gaussian function. A few examples are shown in panel (b). From these fits, we extract the variance of the spatial profile as a function of probe delay. As shown in panel (c), the variance increases rapidly within the first 2 ps. The increase in this range appears linear, with a slope corresponding to a diffusion coefficient of $160 \pm 10$~cm$^2$~s$^{-1}$. At later times, the rate of increase slows down, which may be attributed to exciton localization or additional contributions to the measured signal. At 2 ps, the signal has decayed to about half of its peak value, indicating that a significant portion of the exciton dynamics have already occurred.

We next investigated exciton diffusion along the polar direction by immediately repeating the measurement under identical conditions, except that the probe spot was scanned along the polar axis. The results are shown in Figure~\ref{fig:diffusion}(d–f), yielding a diffusion coefficient of $70 \pm 10$~cm$^2$~s$^{-1}$.

These results reveal a novel exciton transport property of NbOI$_2$. The diffusion coefficient along the nonpolar axis is about one order of magnitude larger than typical values reported for previously studied 2D transition metal dichalcogenides \cite{jap113133702,nanoscale96808,acsp43115,nanoscale64915,acsnano82970,nanoscale79526}, and is comparable to the ambipolar diffusion coefficients reported for 2D cubic boron arsenide \cite{s377433} and carbon nanotube films \cite{b86205417}. It approaches values found in some of the best 2D materials known for superior carrier transport, such as black phosphorus \cite{acsnano96436}. The exciton transport also exhibits anisotropy by a factor of about two. Together with its unique ferroelectricity, these results suggest that NbOI$_2$ could serve as a promising platform for future ultrafast excitonic devices that integrate ferroelectric and anisotropic functionalities. 

The exciton transport in NbOI$_2$ revealed here is related to, but fundamentally different from, charge carrier transport typically studied by electrical techniques. Excitons are neutral quasiparticles; as a result, their scattering by phonons and charged impurities is generally weaker than that of charged carriers, often leading to larger diffusion coefficients. Nevertheless, it is interesting to compare exciton and charge transport in NbOI$_2$. Using the Einstein relation, an effective exciton mobility can be estimated from the measured diffusion coefficient $D_\mathrm{np} = 160$~cm$^2$~s$^{-1}$ via $\mu = {eD}/{k_{\mathrm{B}}T}$, yielding a value on the order of $6 \times 10^3$~cm$^2$~V$^{-1}$~s$^{-1}$, where $e$, $k_{\mathrm{B}}$, and $T = 300$~K are the elementary charge, Boltzmann constant, and temperature, respectively. Interestingly, theoretical studies have predicted room-temperature electron mobilities of the same order of magnitude in monolayer NbOI$_2$ \cite{jpcl1510032,apl128012104}. However, experimentally extracted carrier mobilities are much lower, on the order of $\sim 1$~cm$^2$~V$^{-1}$~s$^{-1}$ \cite{acsnano177170}. The observed anisotropy, with a factor of about two between the nonpolar and polar directions, is also consistent with both theoretical predictions \cite{jpcl1510032,apl128012104} and an experimental estimate based on piezoelectric nanogenerator outputs \cite{ss42300125}.

\section{Conclusions}

In summary, our results show that NbOI$_2$ supports both fast and strongly anisotropic exciton transport, revealing the influence of its low-symmetry ferroelectric lattice on nonequilibrium carrier dynamics. The large diffusion coefficient and directional transport distinguish NbOI$_2$ from many previously studied van der Waals semiconductors and suggest new opportunities for engineering excitonic functionalities using intrinsic polarization. Furthermore, our comparative stability study shows that the combination of inert-environment fabrication, hBN encapsulation, and thermal annealing provides the best long-term stability among the procedures investigated. Together with our prior work on exciton dynamics, these findings provide a more complete picture of the spatiotemporal evolution of photoexcited states in NbOI$_2$ and establish a practical foundation for future studies of ferroelectric excitonic phenomena in this air-sensitive material.

\begin{acknowledgments}
This work was supported by the U.S. Department of Energy, Office of Science, Basic Energy Sciences, under Award No.~DE-SC0020995.
\end{acknowledgments}


%

\end{document}